# Comparison of the COBE DMR and Tenerife Data

C. H. Lineweaver[1,2], S. Hancock [3], G. F. Smoot[1], A. N. Lasenby [3],

R. D. Davies [4], A. J. Banday [5], C. M. Gutiérrez de la Cruz [4], R. A. Watson [6],

R. Rebolo [6]

[1]Lawrence Berkeley Laboratory, Building 50-205, University of California, Berkeley CA 94720.

[2]e-mail: lineweaver@astra.lbl.gov

[3]Mullard Radio Astronomy Observatory, Cavendish Laboratory, Madingley Road, Cambridge CB3 OHE, UK.

[4]University of Manchester, Nuffield Radio Astronomy Laboratories, Jodrell Bank, Macclesfield SK11 9DL, UK.

[5]Universities Space Research Association, NASA/Goddard Space Flight Center, Code 685, Greenbelt, Maryland, 20771.

[6]Instituto de Astrofísica de Canarias, 38200 La Laguna, Tenerife, Spain.

## ABSTRACT

We have compared the Tenerife data with the *COBE* DMR two-year data in the declination $+40°$ region of the sky observed by the Tenerife experiment. Using the Galactic plane signal at $\sim 30$ GHz, we show that the two data sets are cross-calibrated to within 5%. The high Galactic latitude data were investigated for the presence of common structures with the properties of cosmic microwave background (CMB) fluctuations. The most prominent feature in the Tenerife data ($\Delta T \sim 80 \mu$K) is evident in both the 53 and 90 GHz DMR maps and has the Planckian spectrum expected for CMB anisotropy. The cross-correlation function of the Tenerife and DMR scans is indicative of common structure and at zero lag has the value $C(0)^{1/2} = 34^{+13}_{-15} \, \mu$K. The combination of the spatial and spectral information from the two data sets is consistent with the presence of cosmic microwave background anisotropies common to both. The probability that noise could produce the observed agreement is less than 5%.

*Subject headings:* cosmic microwave background — cosmology: observations

## 1. Introduction

The *COBE* DMR has detected structure in the microwave background consistent with fluctuations expected from gravitational instability theories of structure formation (Smoot et al. 1992). The DMR two-year maps contain *rms* temperature fluctuations on 7° scales of $36 \pm 4$ $\mu$K (Banday et al. 1994, Bennett et al. 1994). At lower frequencies, the mountain-based Tenerife experiments (Davies et al. 1992, Watson et al. 1992) operating on $\sim 5°$ scales report an *rms* for the triple beam data of $42 \pm 9$ $\mu$K (Hancock et al. 1994) associated with well-defined CMB structures. A comparison of the DMR and Tenerife data sets can offer a consistency check on the results, the foreground emission and systematic errors of both experiments and provide improved definition of CMB structures.

Ganga et al. (1993) have compared the first year DMR maps with the 170 GHz data from the MIT/FIRS balloon-borne anisotropy experiment (3.8° FWHM). The cross-correlation function of these two data sets indicates the presence of structure common to both. At 170 GHz, the dominant foreground is Galactic dust emission. In this work we extend the comparison to lower frequencies where the dominant foregrounds are expected to be free-free and synchrotron emission. Our comparison combines the spatial and spectral information of the DMR two-year maps with the most recent Tenerife data at declination 40°over the same range in right ascension (R.A.) analyzed by Hancock et al. (1994). In this high Galactic latitude region ($b > 56°$), the Tenerife scans are least contaminated by Galactic emission and the DMR integration time is higher than average, so that a direct comparison of features is possible.

## 2. Methodology

The similar sensitivities and angular scales of the DMR and Tenerife experiments invite a comparison of the observations as a logical extension of the data analysis. A direct comparison of the two data sets is hindered by the limited overlap of the filter functions of the two experiments (Watson et al. 1992) and the small region of the sky ($\approx 4\%$) sampled by the declination 40° Tenerife scans. The different observing strategies, beam profiles, and noise levels of the Tenerife and DMR experiments must be considered. The DMR instrument has observed the full sky at 31.5, 53 and 90 GHz with a beam approximated by a 7.0° FWHM Gaussian (dispersion $\sigma_D = 3.0°$) (Toral et al. 1989)

$$B_D(\mathbf{n}, \mathbf{n}') = \frac{1}{2\pi\sigma_D^2} \exp(-\frac{\theta^2}{2\sigma_D^2}), \tag{1}$$

where $\mathbf{n} = (\alpha, \delta)$ is the position vector of the beam center in right ascension ($\alpha$) and declination ($\delta$), $\mathbf{n}' = (\alpha', \delta')$ is the sky position and $\theta = \cos^{-1}(\mathbf{n} \cdot \mathbf{n}')$. In contrast, the Tenerife experiments observe at frequencies of 10.4, 14.9 and 33 GHz, by drift scanning in right ascension at constant declination and concentrating on a limited sky area. The double-switching with a beam throw of 8.1° is described by the triple beam profile

$$B_T(\mathbf{n}, \mathbf{n}') = \frac{1}{2\pi\sigma_T^2} \left[ \exp(-\frac{\theta^2}{2\sigma_T^2}) - \frac{1}{2} \left( \exp(-\frac{\theta_e^2}{2\sigma_T^2}) + \exp(-\frac{\theta_w^2}{2\sigma_T^2}) \right) \right], \tag{2}$$

where $\theta_e = \cos^{-1}(\mathbf{n_e} \cdot \mathbf{n}')$, $\theta_w = \cos^{-1}(\mathbf{n_w} \cdot \mathbf{n}')$, $\mathbf{n_e}$ and $\mathbf{n_w}$ are direction vectors $\simeq 8.1° \times \sec\delta$ degrees in right ascension East and West of $\mathbf{n}$ and the beam dispersion is $\sigma_T = 2.2°$. Given a true sky brightness temperature distribution $T(\mathbf{n}, \nu)$, the DMR and Tenerife data can be represented by

$$T_D(\mathbf{n}, \nu) = \int\int d\Omega_{\mathbf{n}'} T(\mathbf{n}', \nu) B_D(\mathbf{n}, \mathbf{n}') + N_D(\mathbf{n}, \nu) \tag{3}$$

$$T_T(\mathbf{n}, \nu) = \int\int d\Omega_{\mathbf{n}'} T(\mathbf{n}', \nu) B_T(\mathbf{n}, \mathbf{n}') + N_T(\mathbf{n}, \nu), \tag{4}$$

where $N_D(\mathbf{n}, \nu)$ and $N_T(\mathbf{n}, \nu)$ are the position and frequency dependent noise in the DMR maps and the Tenerife scans respectively. It is clear from equations (3) and (4) that the sky structure in the two data sets has been smoothed by different amounts ($\sigma_D > \sigma_T$), and in the case of Tenerife, the double switching has removed some of the large scale power sampled by the DMR instrument. We wish to make the sky structures in one data set as directly comparable as possible with those in the other. Since the Tenerife data are one-dimensional scans, they do not contain the two-dimensional information necessary to allow a two-dimensional smoothing to the lower DMR resolution. To compare the DMR data to the Tenerife data, we run the Tenerife triple beam over the DMR maps at declination 40° to produce double-differenced DMR scans defined by

$$T_{D,scan}(\mathbf{n}, \nu) = b \int d\Omega_{\mathbf{n'}} T_D(\mathbf{n'}, \nu) B_T(\mathbf{n}, \mathbf{n'}), \tag{5}$$

where $\mathbf{n} = (\alpha, 40°)$. Structure in the Tenerife scans has been smoothed by the 5.1° Tenerife beam and, with this method, the same structure in the DMR scans has been smoothed by the 7° DMR beam and the 5.1° Tenerife beam. The extra 5.1° smoothing of the DMR maps to a net resolution of 8.7° has the positive effect of reducing the noise contribution to the DMR scans, but broadens features as seen in Figure 1. A boost factor $b$ is introduced to recover the amplitude of the smoothed DMR scans. We have experimented with other methods of comparing the data including comparing the double differenced Tenerife scans to unsmoothed DMR maps and to unsmoothed but double differenced DMR scans. Monte Carlo tests with appropriate noise levels and a variety of signals indicated that none of these methods was significantly better than our chosen method.

## 3. Cross-Calibration on the Galactic Plane

As a first step in the comparison, we cross-calibrate the DMR 31.5 GHz data with the Tenerife 33 GHz data using the largest signal present in both data sets, namely the Cygnus region of the Galactic plane. The declination 40° Tenerife drift scans pass through the emission peak centered at $\mathbf{n}_{Cyg} \approx (308°, 40°)$, producing the characteristic triple beam profile (Figure 1a). The corresponding DMR scan produced from the two-year 31.5 GHz $(A + B)/2$ map is also shown. The DMR scan has a smaller amplitude and is not directly comparable to the Tenerife scan because of the additional DMR beam smoothing and the difference in observing frequency.

We model the Cygnus source as a Gaussian free-free emission source of amplitude $A$ and dispersion $\sigma_{Cyg}$

$$G(\mathbf{n}', \nu) = A \exp\left(-\frac{\theta^2}{2\sigma_{Cyg}^2}\right) \left(\frac{\nu(\text{GHz})}{33\,\text{GHz}}\right)^{\beta}, \qquad (6)$$

where $\theta = \cos^{-1}(\mathbf{n}_{Cyg} \cdot \mathbf{n}')$. Fitting this model to the Tenerife 33 GHz data using equation (4) with $T(\mathbf{n}', \nu) = G(\mathbf{n}', \nu)$ and $N_T(\mathbf{n}, \nu) = 0$, one obtains the dashed line in Figure 1a, corresponding to $A = 9.5$ mK and $\sigma_{Cyg} = 3.0°$. We scale the model from 33 GHz to 31.5 GHz using the canonical free-free spectral index $\beta = -2.1$. Measurements of $\beta$ in the Cygnus region from the Tenerife and DMR scans are consistent with this choice: $\beta = -2.0 \pm 0.1$ from the 15 vs 33 GHz data and $\beta = -2.1 \pm 0.2$ from the 31.5 vs 53 GHz data. With the above values for $A$, $\sigma_{Cyg}$ and $\beta$, we use $G(\mathbf{n}', \nu)$ as the true sky in equation (3) with $N_D(\mathbf{n}, \nu) = 0$. Equation (5) then produces the DMR model scans seen in Figure 1a. The agreement of the DMR model scan with the DMR data scan is thus a measure of the agreement of the two experiments. The amplitudes of the DMR data and model agree to within 5%.

The reduced amplitude of the DMR 31.5 GHz scan as compared to the Tenerife 33 GHz scan is a result of the extra smoothing in the DMR instrument beams. To a good approximation one can restore the amplitude with a boost factor $b$ based on the ratio of the peak of the signals in the two model scans. We find $b \approx 2$ and apply this to the DMR data to obtain the DMR scan points in Figure 1b. The emission peak is restored to the expected frequency dependent amplitude and the shape of the feature is approximately recovered. We conclude that our comparison method is effective; that the DMR and Tenerife experiments consistently observe known structures, and that the DMR data is cross-calibrated to better than 5% with the Tenerife data.

In general the form of $T(\mathbf{n})$ is not known and a generic boost factor cannot be derived since the degree of boosting necessary to restore a given feature is a function of the two-dimensional form of the feature. However, making plausible assumptions about the spatial variation of features in the scans, one can derive an approximate boost factor with which to restore such features. We derive boost factors based on the hypotheses that the spatial variation of the structures can be described by CMB fluctuations or by Galactic synchrotron/free-free structures. A boost factor $b = 1.5 \pm 0.3$ recovers the amplitude of the high peaks for a broad range of possible sources. This value is lower than the $b = 2.0$ value obtained for the Galactic plane because the latter is an isolated high signal to noise peak, whereas more realistically we are dealing with a superposition of peaks and valleys of comparable amplitude. The required level of boosting is not significantly affected by the precise form of the features; we use boosts of 1.2, 1.5 and 1.8 and find that our conclusions do not strongly depend on the boost value.

## 4. Scan Comparison

Having tested our comparison method on the Galactic plane ($\sim 5$ mK level), we investigate the high Galactic latitude region, for which Hancock et al. (1994) report the existence of CMB features in the Tenerife data ($\lesssim 0.1$ mK level). The separate scans from each of the six frequencies are presented in Figure 2; the DMR scans are from equation (5) with $b = 1.5$. In the absence of sky structure, the data points in the Tenerife scans are independent. This is not the case for the DMR scans which, by virtue of the comparison method, contain correlated noise structures. The most prominent feature in the Tenerife data is a $\sim 3\sigma$ signal at $170° \lesssim$ R.A. $\lesssim 210°$ in the Tenerife 15 and 33 GHz scans. This feature is evident in both the 53 and 90 GHz DMR scans but not in the less sensitive 31.5

GHz scan (note the difference in scale in the top panels). The large feature in the 53 GHz scan centered at R.A. 220° is not in the other DMR channels and may be partially due to a noise spike overlying the lower amplitude feature indicated by the Tenerife scans.

The well correlated scans over the R.A. region $170° - 210°$ (Galactic latitude $b > 73°$) suggest the presence of common structure which may be CMB anisotropy. The possibility that the signals originate from foreground point radio sources is excluded by the relative insensitivity of the higher frequency data to such sources of emission (e.g. 1 Jy contributes $3~\mu$K to the 33 GHz scan and 0.6 $\mu$K to the 53 GHz scan) and by the absence of strong sources in the sky area under observation (Condon et al. 1986, 1989 and Kuhr et al. 1981). The potential Galactic components are synchrotron, free-free and dust emission and the observed antenna temperatures have the characteristic frequency dependence (see e.g. Bennett et al. 1992)

$$\frac{T_A(\nu_1)}{T_A(\nu_2)} = \left(\frac{\nu_1}{\nu_2}\right)^{\beta}. \tag{7}$$

Assuming the values $\beta_{sync} = -2.75$, $\beta_{ff} = -2.1$ and $\beta_{dust} = 1.5$ for the spectral indices, a visual inspection of Figure 2 indicates that no Galactic foreground is present; there are no spatially correlated bumps which change by the ratios 420:1 (synchrotron), 93:1 (free-free) and 1:25 (dust) between 10 and 90 GHz. To confirm this impression, a separate $\chi^2$ was calculated for each suspected component of Galactic emission and also for a Planckian spectrum. We assume that the structure is best modelled by a weighted combination of the 10, 15, 33, 31, 53 and 90 GHz scans at an average frequency $\overline{\nu}$ where the combination is formed so as to be consistent with the spectrum under consideration

$$\overline{\Delta T}_{A,i}(\overline{\nu}) = \frac{\sum_{j=1}^{nfreqs=6} \Delta T_{A,i}(\nu_j) r(\nu_j) w_i(\nu_j)}{\sum_{j=1}^{nfreqs=6} w_i(\nu_j)}, \tag{8}$$

for points $i$ in R.A. with weight $w_i(\nu_j) = 1/(r(\nu_j)\sigma_i(\nu_j))^2$ and temperature $\Delta T_{A,i}(\nu_j)$ observed at a frequency $\nu_j$. The conversion factors $r(\nu_j) = 1/R(\nu_j)$ are defined according to the spectrum under consideration; for Galactic foregrounds

$$R(\nu_j) = \left(\frac{\nu_j}{\overline{\nu}}\right)^{\beta},\tag{9}$$

and for a blackbody

$$R(\nu_j) = \frac{f(\nu_j)}{f(\overline{\nu})},\tag{10}$$

where $f(\nu) = x^2 e^x/(e^x - 1)^2$ and $x = h\nu/kT_{cmb}$. We form a $\chi^2$ for each of the candidate spectra:

$$\chi^2 = \sum_{i=170}^{210} \sum_{j=1}^{nfreqs=6} \frac{(\Delta T_{A,i}(\nu_j) - \overline{\Delta T}_{A,i} R(\nu_j))^2}{\sigma_i^2(\nu_j) + (\sigma_{\overline{\Delta T}_{A,i}} R(\nu_j))^2}.\tag{11}$$

The fit to a blackbody source yields $\chi^2 = 32$ for 55 degrees of freedom, with any single Galactic component being ruled out at $\gtrsim 90\%$ confidence level. This conclusion is independent of the boost parameter in the range tested; $1.2 \leq b \leq 1.8$. The analysis favors a Planck spectrum consistent with CMB fluctuations, but does not rule out a combination of foreground contaminants mimicking a blackbody source. The analysis of Galactic emission by Bennett et al. (1992) suggests that the latter is unlikely and this is supported by a multi-component fit to the Galactic foregrounds (Hancock et al. in preparation).

In order to obtain optimum sensitivity to CMB structures, combined scans for each experiment are made by weighted addition of the separate scans after conversion to thermodynamic temperature. It is likely that the lowest frequency (10 GHz) scan contains an appreciable contribution from Galactic synchrotron or free-free emission and is excluded, whereas the dust emission, even in the highest frequency 90 GHz scan, is not expected to be significant (Bennett et al. 1992). The similarity of the combined scans in the lowest panels of Figure 2 provides evidence that the DMR and Tenerife instruments are observing the same CMB features. The apparent agreement of the DMR combined scan with the Tenerife combined scan can be quantified with the cross-correlation function

$$C(\alpha) = \frac{\sum_{i,j} w_{D,i} w_{T,j} \Delta T_{D,i} \Delta T_{T,j}}{\sum_{i,j} w_{D,i} w_{T,j}},\tag{12}$$

where $w_{D,i}, \Delta T_{D,i}$ and $w_{T,j}, \Delta T_{T,j}$ are the weights and double-differenced temperatures of the DMR and Tenerife combined scans respectively. The sums are over all pairs $(i, j)$ with separation angle $\alpha_{ij}$ within half a bin width of $\alpha$. In Figure 3, the data points and error bars represent the cross-correlation function of the DMR combined scan with the Tenerife combined scan. The error bars are the 68% confidence levels from 1000 Monte Carlo simulations of the cross-correlation function obtained from realizations of the data points plus Gaussian noise. Also shown in Figure 3 are dotted curves denoting the one-sigma confidence bounds for a model Harrison-Zel'dovich sky normalized to $Q_{rms-PS} = 26$ $\mu$K (Hancock et al. 1994). The distinctive profile of the observed cross-correlation function implies the presence of structure common to the Tenerife and DMR combined data scans, with the properties expected of a Harrison-Zel'dovich sky with this normalization.

It is possible that the agreement between the DMR and Tenerife scans is simply due to spurious correlations with noise features. We quantify the likelihood of such an occurrence by calculating the cross-correlation function between the combined scans and noise realizations. The light-grey shaded band in Figure 3 is the 68% confidence region derived from 1000 DMR noise simulations cross-correlated with the Tenerife combined scan. The first data point, $C(0)$, lies just above the 90% confidence level from the noise simulations. Given the twin-tailed nature of the distribution, there is thus less than a 5% chance that such a high $C(0)$ could result from noise features in the DMR data correlating with features in the Tenerife data. For the cases of Tenerife noise correlating with DMR data and Tenerife noise correlating with DMR noise the confidence intervals are shown in Figure 3 as a dark grey band and dashed lines respectively. The corresponding probabilities are both $< 0.1\%$. These probabilities are robust to the boost factor, which scales both the data and the noise realizations by the same amount and simply affects the overall normalization. Taking into account the uncertainty in the boost value gives $\sqrt{C(0)} = 34^{+13}_{-15}$ $\mu$K for $1.2 \leq b \leq 1.8$, which is compatible with the value of $\sqrt{C(0)} = 35^{+8}_{-11}$ $\mu$K obtained from the Tenerife data alone. We have also calculated a related statistic, Pearson's linear correlation coefficient $r$. We obtain $r = +0.56$ with a probability of 4% that a correlation coefficient this high or higher would result from the correlation of the Tenerife data with DMR noise. For the case of both data sets being only noise, this probability falls to 0.6%.

The scan with optimum sensitivity to CMB structures is the weighted addition of the Tenerife and DMR combined scans. Figure 4 shows the $(A + B)/2$ and $(A - B)/2$ combinations, for which $\chi^2$ values of 70.0 and 15.2 result for 16 degrees of freedom. Significant structure is present in the $(A + B)/2$ scan, while the $(A - B)/2$ scan is in agreement with a random noise distribution. The *rms* amplitude of the sky signal can be extracted from these two combined scans using $\sigma_{sky} = (\sigma^2_{(A+B)/2} - \sigma^2_{(A-B)/2})^{1/2}$. We obtain $\sigma_{sky} = 37^{+5}_{-8}$ $\mu$K, in agreement with $\sigma_{sky} = 38^{+6}_{-10}$ $\mu$K derived from the 4° binned 15 and 33 GHz scans. This signal is associated with the well-defined structures clearly visible in the combined scan and given the frequency coverage (15-90 GHz) is most consistent with a primary origin in CMB fluctuations.

## 5. Discussion and Conclusions

The Tenerife and DMR data sets are the result of independent observations and by comparing them we significantly reduce the uncertainties associated with systematic errors and foreground contamination. The Galactic plane is observed at the expected amplitude, and the calibration of the two experiments agrees to within 5%. The latter is of particular relevance to using the data to constrain the overall normalization and slope of the primordial power spectrum (Hancock et al. 1994, Figure 3). In the high Galactic latitude region, the principle structures in the two data sets are similar over the frequency range 15-90 GHz. A single component Galactic origin for the structures is rejected at $\gtrsim 90\%$ confidence level. The cross-correlation function between the two data sets converted to thermodynamic temperature implies common spatial and amplitude characteristics for the structures, supporting the existence of common CMB features with only a 5% probability that noise could produce the observed agreement. Furthermore, the cross-correlation function is in agreement with the normalization of $Q_{rms-PS} = 26 \pm 6$ $\mu$K deduced by Hancock et al. (1994) for sky fluctuations described by a Harrison-Zel'dovich spectrum, thus explicitly demonstrating the consistency of this locally large normalization with the overall sky normalization of $Q_{rms-PS} \approx 20\mu$K (Górski et al. 1994, Banday et al. 1994).

The inadequacies of this comparison lie in the boost factor approximation and the reduced power in the DMR scans due to the Tenerife beam filtering. The latter has been addressed by appropriate noise simulations, but the former is an inherent limitation in the comparison method, and in fact degrades our knowledge of the signal level as compared to using the Tenerife data alone. The significance of this comparison lies not in the improvement in the signal to noise of given CMB features, but in the extension of the frequency range over which identifiable features are observed. This comparison is restricted by the low signal to noise of the DMR data, and the one-dimensional nature of the Tenerife scans. Continuing observations with the Tenerife instruments promise to address the latter, while the DMR 4-year data has been collected and is in the process of being reduced. The improvements in both data sets will provide an opportunity to refine the comparison presented here.

The National Aeronautics and Space Administration/Goddard Space Flight Center (NASA/GSFC) is responsible for the design, development, and operation of the *COBE* satellite. The Tenerife experiments are supported by the UK PPARC, and the EU and Spanish DGICYT Science Programmes. C. H. L. acknowledges helpful suggestions from Al Kogut and a NASA GSRP grant. S. H. acknowledges a research fellowship at St. John's College, Cambridge University. We thank all our colleagues involved in the Tenerife and DMR projects whose efforts have made this work possible.

Captions

Figure 1

Cross-calibration of the 31 GHz and 33 GHz data sets, assuming a Gaussian model for the Cygnus region of the Galactic plane. In a) the dashed line shows the best fit model to the 33 GHz data after convolution with the triple beam of the Tenerife experiment. The solid line results from scaling the best fit model to 31 GHz and producing a DMR scan (equation (5)). The agreement of the DMR model scan with the DMR data scan is a measure of the agreement between the results of both experiments. In b) we restore the DMR scan amplitude with a factor $b = 2.0$ to compensate for the additional smoothing of the DMR scans.

Figure 2

The Tenerife and DMR scans at declination 40° at all six frequencies, binned at 4° intervals in R.A. The vertical axes are in antenna temperature, except for the bottom panels which are in thermodynamic temperature. Note the different scale for the least sensitive channels in the top panels. The error bars on the DMR scans were obtained from 1000 simulations of noise added to the DMR maps. The bold lines in the bottom panels correspond to the weighted addition of the scans above (the 10 GHz channel may contain an appreciable Galactic signal and is excluded from the additions). For comparison, the dashed line in g) is the bold line in h) and vice-versa.

Figure 3

Cross-correlation function of the DMR and Tenerife combined scans at declination 40°. The 68% error bars on the data points are obtained from correlating random realizations of both combined scans. The bands are the 68% confidence levels for the cross-correlation of the Tenerife combined scan with the DMR noise (light grey), the Tenerife noise with the DMR combined scan (dark grey), and the Tenerife noise with DMR noise (dashed lines).

The dotted lines are the 68% confidence intervals for a Harrison-Zel'dovich spectrum with $Q_{rms-PS}$ =26 $\mu$K.

Figure 4

The $(A + B)/2$ and $(A - B)/2$ scans obtained from the weighted combination of the 15, 31, 33, 53 and 90 GHz scans using a Planckian spectral form. The vertical axis is the thermodynamic temperature. In both experiments, channels A and B are independent data sets (Smoot et al. 1990, Hancock et al. 1994).

## Galaxy Calibration

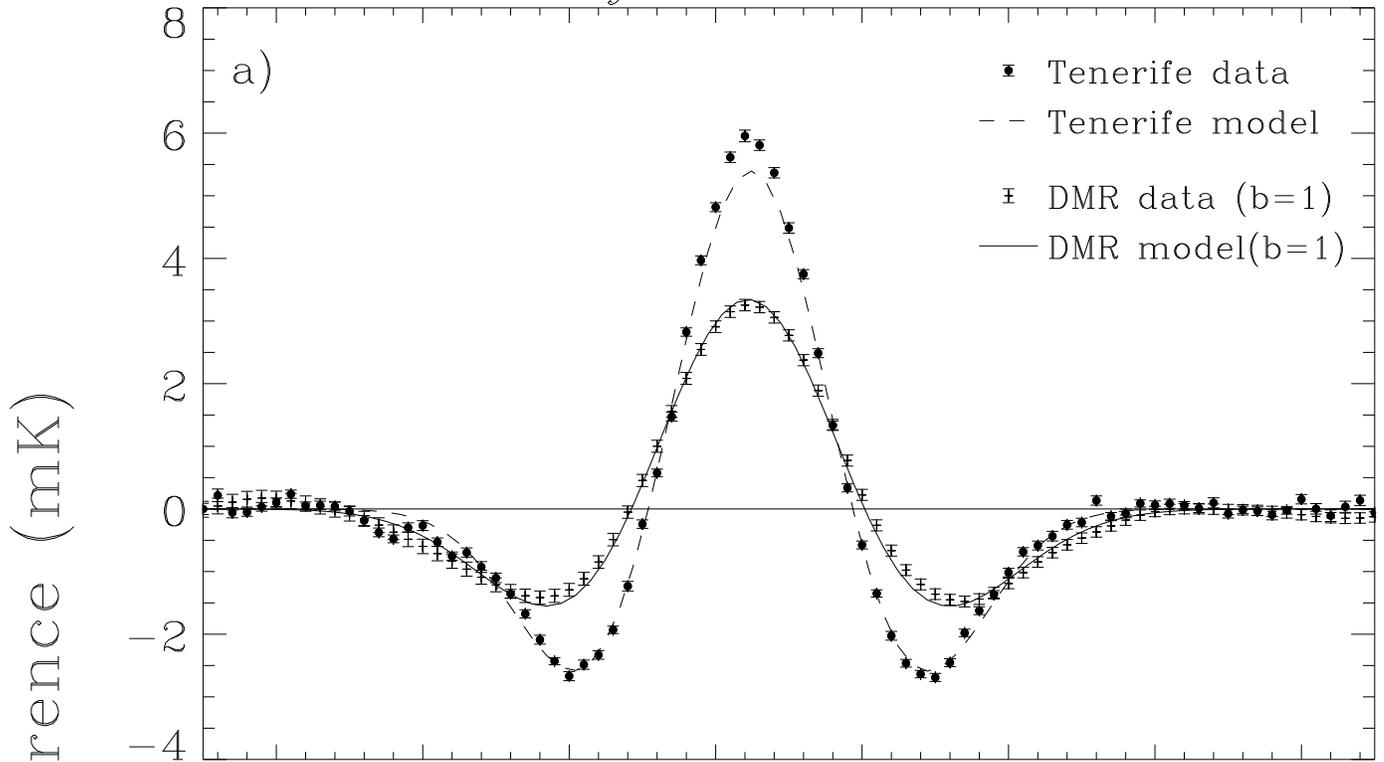

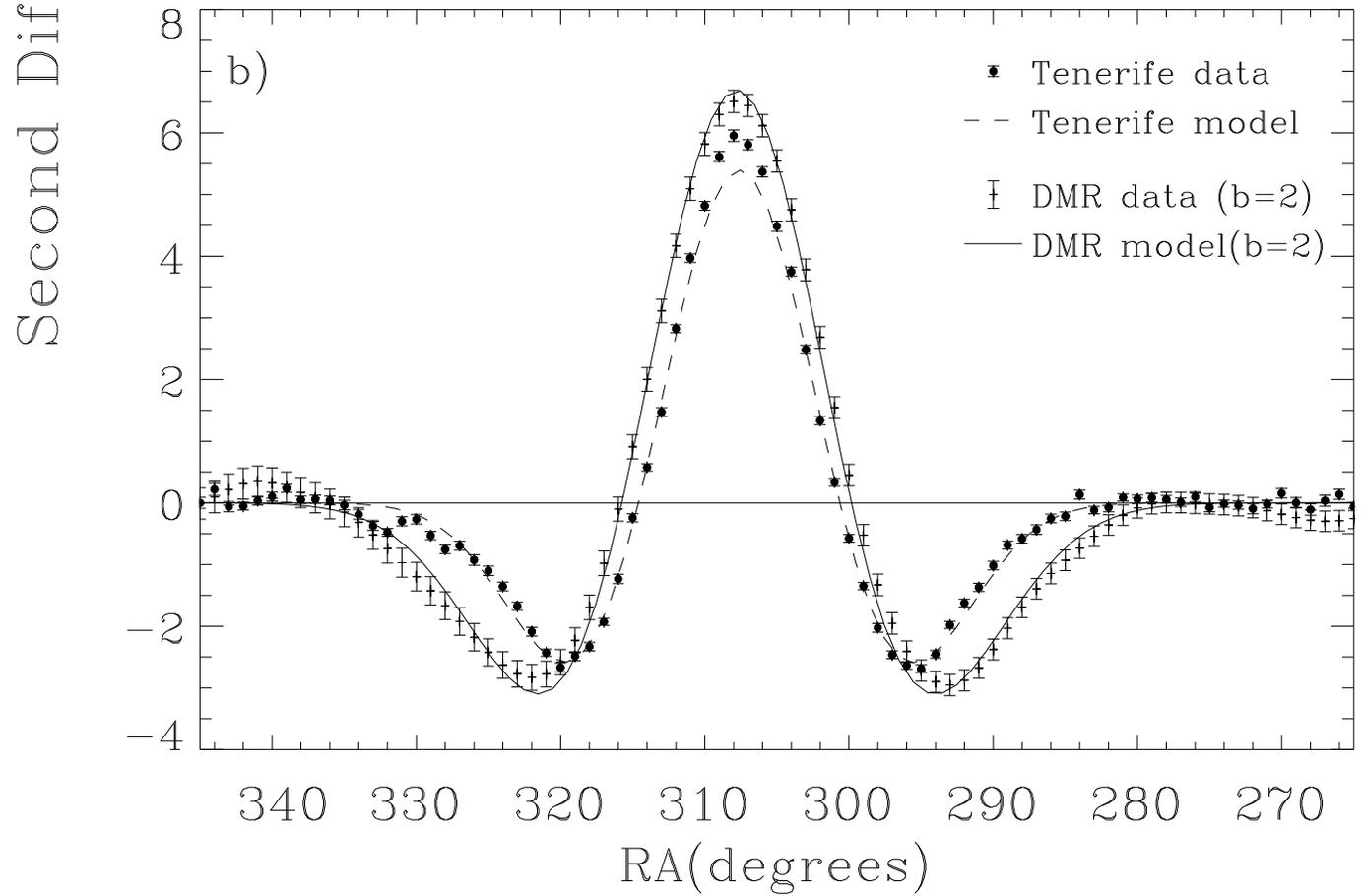

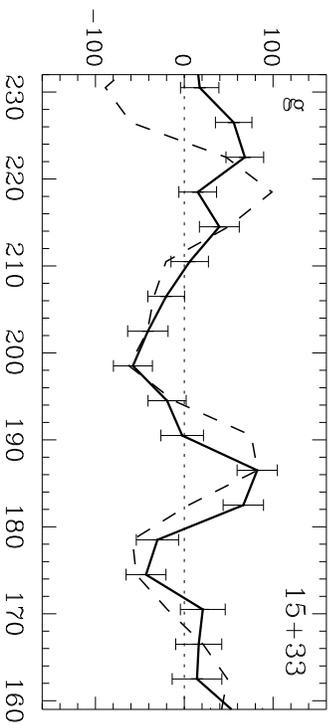
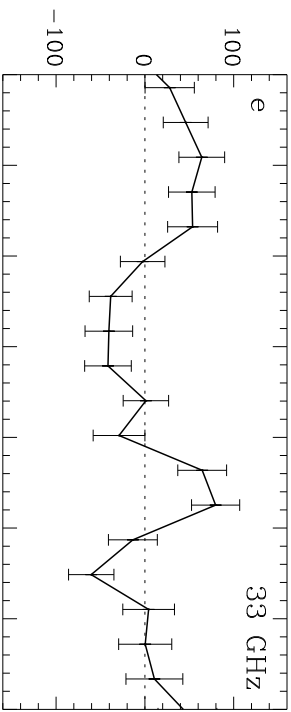
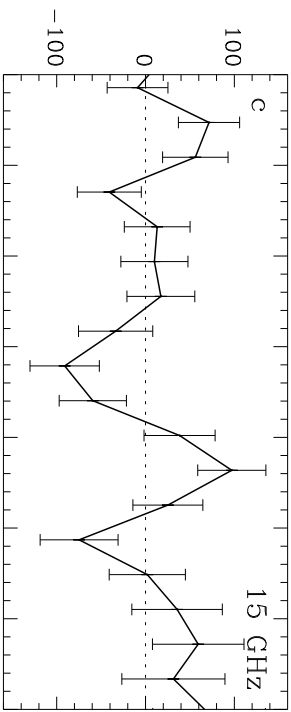
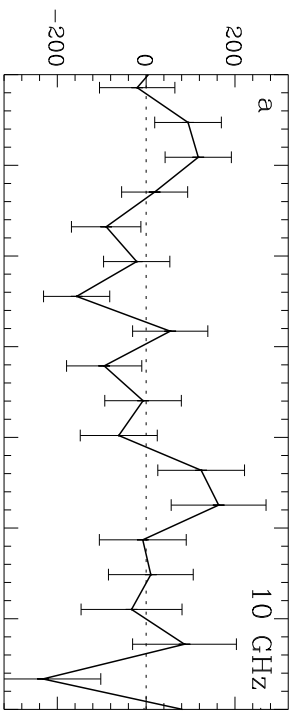
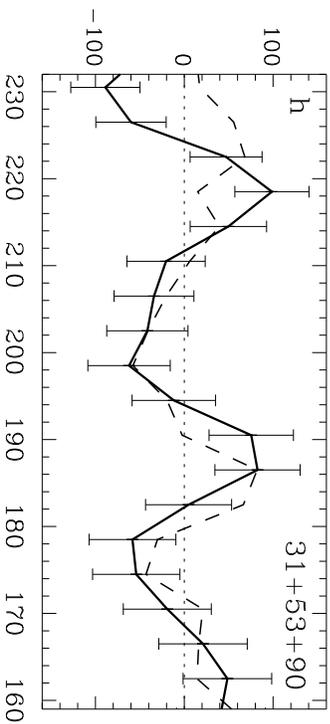
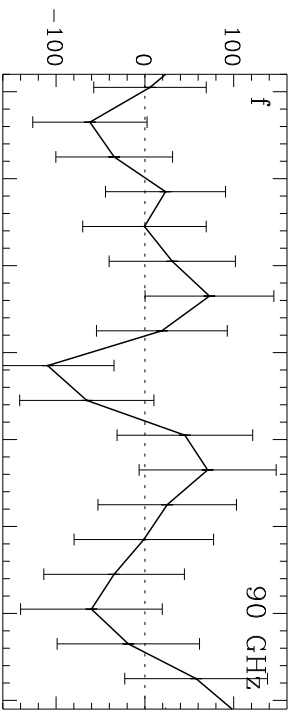
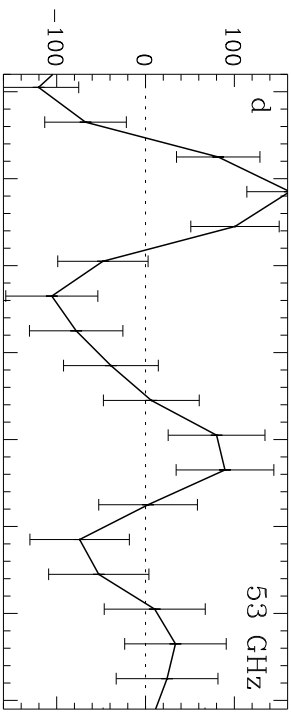
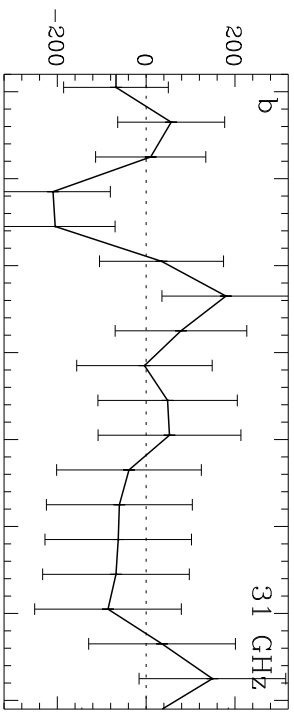

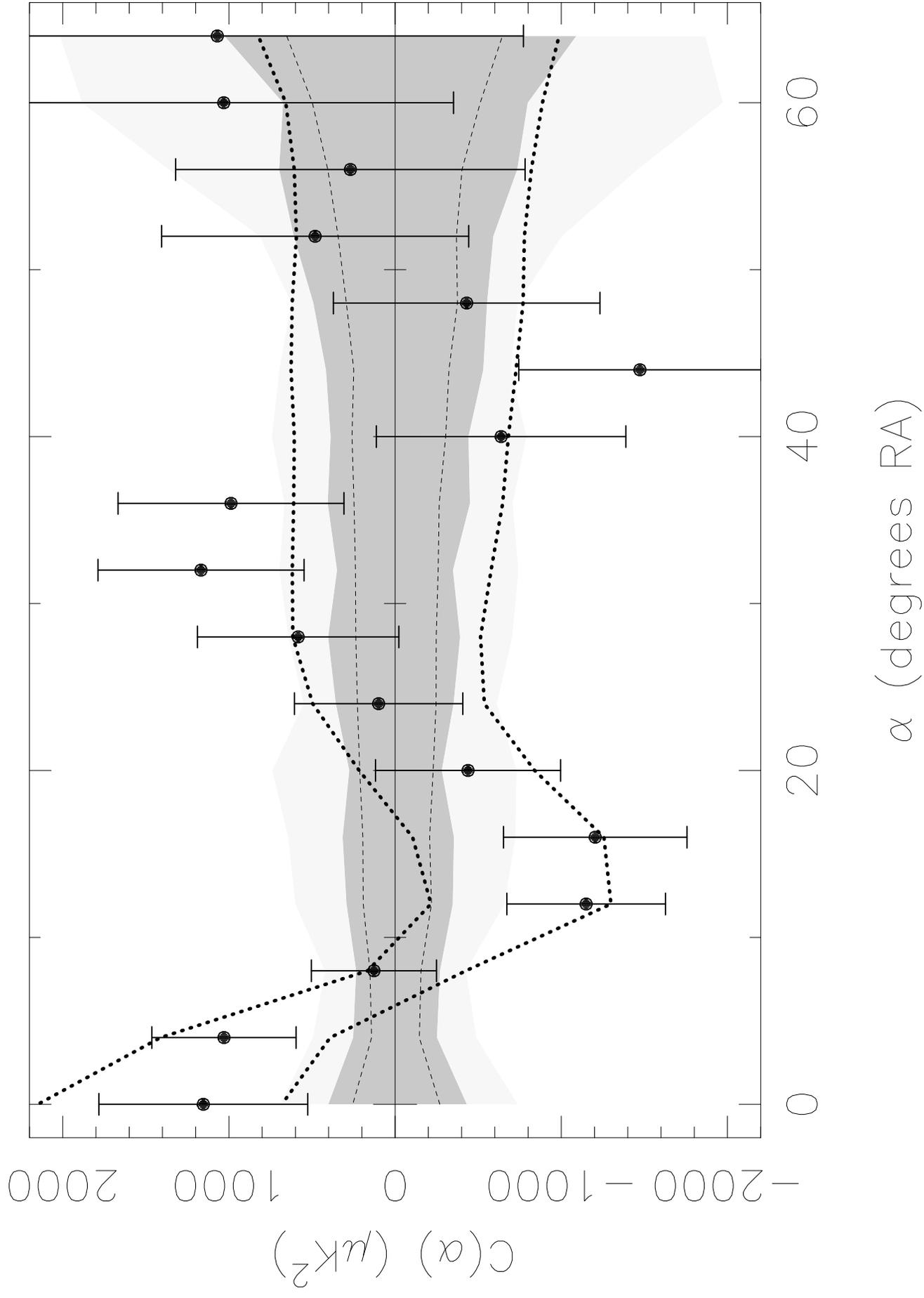

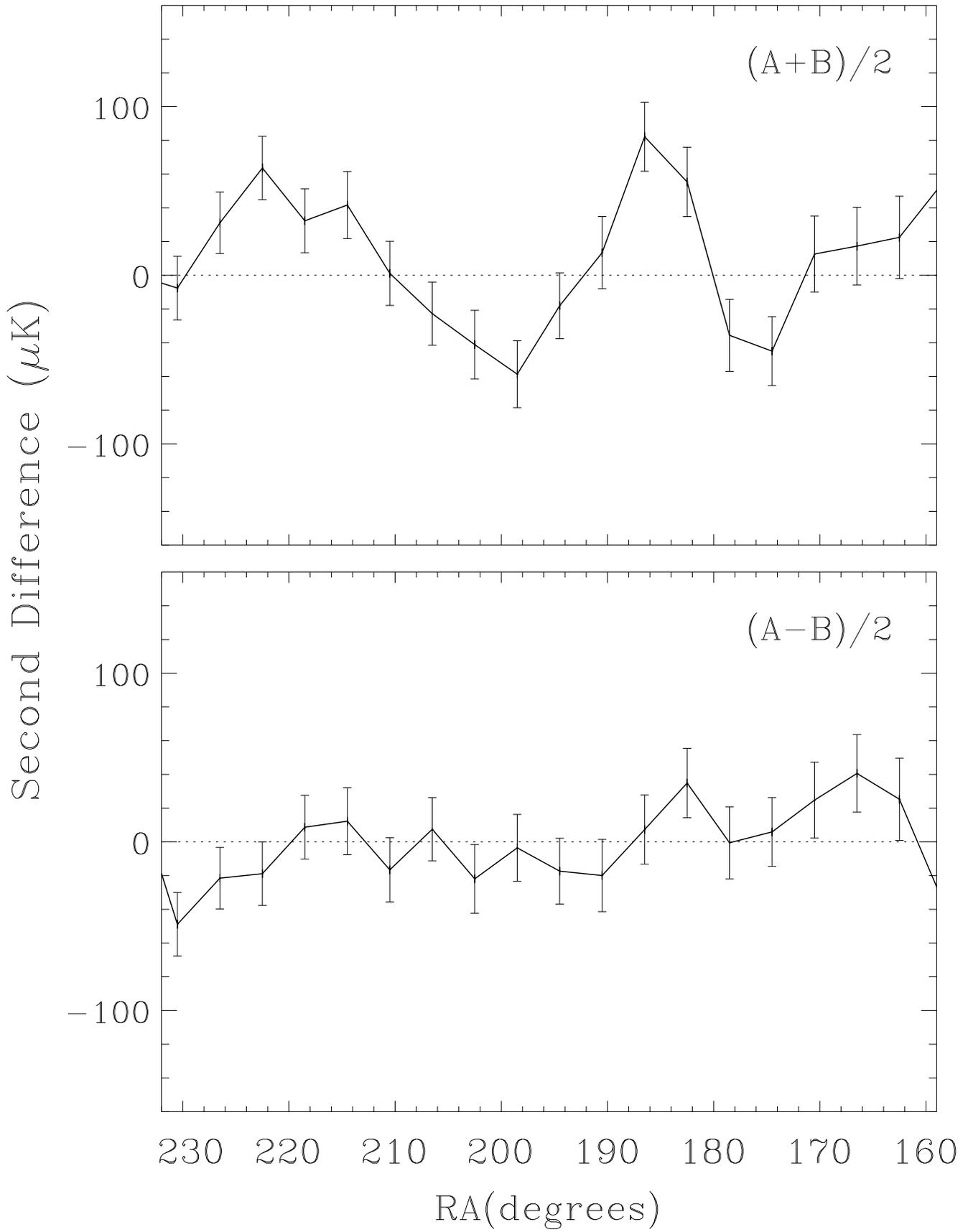